# Numerical Investigation of Unsteady Aerodynamic Effects on Thick Flatback Airfoils

G.S.T.A. Bangga, Th. Lutz, E. Krämer
Institute of Aerodynamics and Gas Dynamics (IAG), University of Stuttgart, Germany
bangga@iag.uni-stuttgart.de

**Summary**

The unsteady characteristics of the flow over thick flatback airfoils have been investigated by means of CFD calculations. Sandia airfoils which have 35% maximum thickness with three different trailing edge thicknesses were selected. The calculations provided good results compared with available experimental data with regard to the lift curve and the impact of trailing edge thickness. Unsteady CFD simulations revealed that the Strouhal number is found to be independent of the lift coefficient before stall and increases with the trailing edge. The present work shows the dependency of the Strouhal number and the wake development on the trailing edge thickness. A recommendation of the Strouhal number definition is given for flatback airfoils by considering the trailing edge separation at low angle of attack. The detailed unsteady characteristics of thick flatback airfoils are discussed more in the present paper.

## 1. Introduction

As the needs of wind energy increase, the size of wind turbines also increases significantly nowadays. Accordingly, the blade size increases and therefore the need of higher inertial moment at the blade root becomes increasingly important. The inboard sections of the blade (about r/R < 35%) normally consist of very thick airfoils. However, due to the excessive thickness, the aerodynamic performance of the sectional airfoil is inferior compared with the thinner airfoil section. This has been reported in Ref. [1] for airfoils with maximum thickness (t/c) more than 25%. The shape of these airfoils provides very high adverse pressure gradients (APG), and hence promote stronger separation. Moreover, the leading edge of airfoils cannot be avoided from contamination [2], and this leads to premature laminar-turbulent transition of the flow over the profile. Measurements from [3] and [4] for thick airfoils under tripped condition have shown an earlier stall with a significant reduction of the maximum lift coefficient ($c_{l,max}$). This becomes a challenge for blade designers to reduce the sensitivity of the airfoil towards surface roughness near the leading edge.

Flatback airfoils are introduced to increase structural and aerodynamic performances at the inner region of wind turbine blades. Aerodynamically, the flatback reduces the adverse pressure gradient, which in turn increases achievable sectional maximum lift coefficient and lift curve slope, and reduces sensitivity of the lift characteristics to surface soiling [3]. When the blade root operates at high angles of attack (AoAs) and stalls, centrifugal force transports the separated boundary layer from the root towards the middle region of the blade and it has adverse effects on the performance of the blade [5]. As the centrifugal force plays a significant role only for the separated flow, the underlying phenomena are expected to be less pronounced for the blade with flatback trailing edge (TE). However, due to the increase of trailing edge thickness, the flow over the blade is more unsteady due to the trailing edge separation. The blunt trailing edge generates counter-rotating vortices detaching periodically from the upper and lower edge of the base even at low angle of attack. This evokes the fluctuation of forces acting on the airfoil and there are only limited studies available on this matter. Therefore, this paper is intended to gain more information on the unsteady characteristics of flatback airfoils by means of computational fluid dynamics (CFD).

## 2. Computational Method

The 35% maximum thicknesses Sandia airfoils FB-3500-0005, -0875 and -1750 with three different trailing edge thicknesses respectively 0.5%, 8.75%, and 17.5% from Ref. [3] were selected in the investigation as shown in Fig. 1. The available measurement data [3] have been taken in the mid-span plane of the two-dimensional (2D) airfoil model, therefore, it is reasonable to use 2D geometrical configuration for present calculations. The mesh was C-type and constructed using IGG grid generator with automesh script developed at IAG. The mesh size is of the order of $2 \times 10^5$ with 480 nodes on the airfoil surface. To capture the viscous flow within the laminar sub-layer accurately, the mesh was refined near the airfoil as shown in Fig. 2 and the height of the first layer of the cells was set to meet the non-dimensional wall distance $y^+ < 1$ [3,6-10].

The CFD code FLOWer was used for the unsteady numerical calculations. The (URANS) SST $k$-$\omega$ turbulence model was employed to model the turbulence. To accelerate the convergence, multigrid level 3 was utilized. Dual time-stepping was applied to obtain second-order accuracy in time. Cell vertex discretization scheme was chosen as it provides high accuracy and high validation level [7]. All simulations were carried out according to experimental data from Ref. [3] at standard sea level conditions. The Reynolds number is 330,000. Tripped conditions are modelled by fixing the transition in the calculations at 2% and 5% on the suction and pressure side, respectively. The calculations were carried out using LAKI (NEC Cluster) from HLRS and the parallelization was done using the message passing interface (MPI) with 7 processors in each calculation.

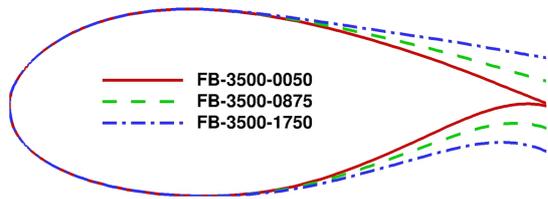

Fig. 1 Sandia airfoils with three different trailing edge thicknesses [3].

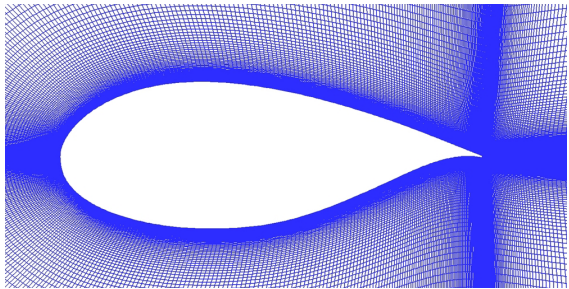

Fig. 2 Detailed mesh near the FB-3500-0050 airfoil.

## 3. Results and Discussion

The unsteady characteristics of flatback airfoils with three different trailing edge thicknesses have been investigated numerically. Grid and timestep studies have been conducted to show that the solutions are independent of the spatial and temporal resolutions, and all presented results are based on the selected grid with timestep value lower than 1% of the convection time of a single fluid particle passing the airfoil.

3.1 Validation of the Results
The computational results in term of lift polars for the airfoils with three different trailing edge thicknesses are validated against available experimental data [3], see Fig. 3. Present calculations show a good agreement with the measurements. The value of lift coefficient ($c_\ell$) at pre-stall conditions as well as the location of maximum $c_\ell$ was predicted accurately though the measured lift drop for the 17.5% trailing edge thickness airfoil was not captured and the undershoot for the 0.5% trailing edge thickness airfoil was over-predicted. The Overflow results from Ref. [3] were included for comparison. The CFD calculations of airfoils under stall condition at high AoA remain challenging since they exhibit highly separated flow and being the subject of 3D phenomena [6,8-11]. The 2D CFD models are expected to struggle predicting this type of flow (moreover, the turbulence is not resolved but is modelled). Therefore, the choice of the turbulence model can add to the variability of the predicted maximum lift [12].

The airfoil with 0.5% trailing edge thickness shows the presence of stall at very low AoA (about -2°). The reason behind this phenomenon is the occurrence of high adverse pressure gradient on both pressure and suction side of the airfoil. The value of minimum pressure coefficient ($c_p$) reduces as the AoA increases from -2° to 2°. The pressure side contribution increases and reaches its maximum as the airfoil reaches the minimum value of $c_\ell$. It can be seen in Fig. 4 that the pressure side has lower $c_p$ value than the suction side for x/c > 0.4 at AoA = 2°, which in turn produces negative $c_\ell$ for airfoil with 0.5% trailing edge thickness. The increase of trailing edge thickness reduces the adverse pressure gradient, increase the minimum value of $c_p$, hence the $c_\ell$ increases, but the drag coefficient ($c_d$) increases due to the trailing edge separation. However, near the hub, the value of drag contributes only a small amount to the torque generation as it is generated mostly due to the lift force. This comes from the fact that the ratio of lift to drag ($c_\ell /c_d$) increases with the trailing edge for airfoils under tripped conditions, as also reported in [3]. Therefore, it is expected that the drag increase with the trailing edge is less important and the flatback airfoils are still feasible to be implemented in the large wind turbine blades.

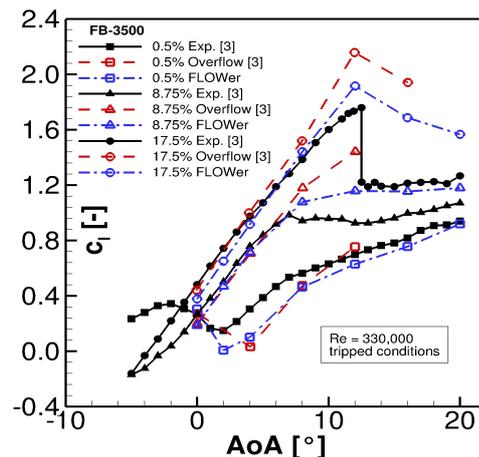

Fig. 3 Comparison of lift coefficient between CFD calculations and experimental data.

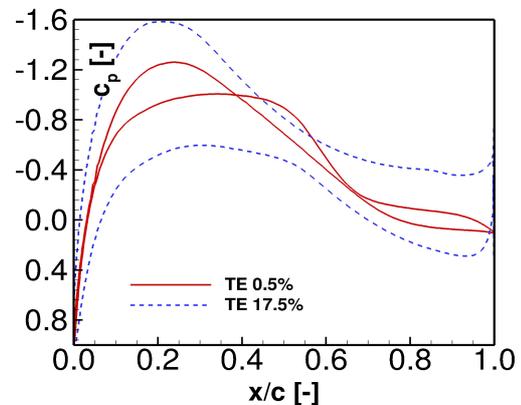

Fig. 4 Averaged cp distribution of flatback airfoils FB-3500-0050 and -1750 at AoA 2°.

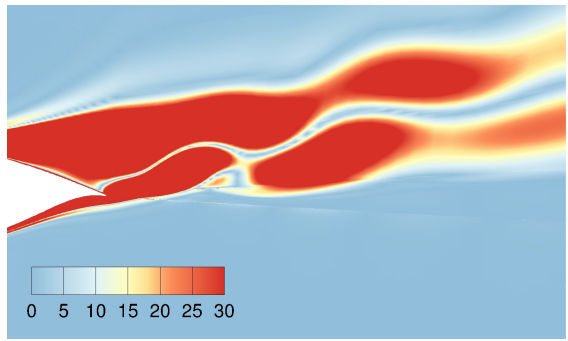

(a)

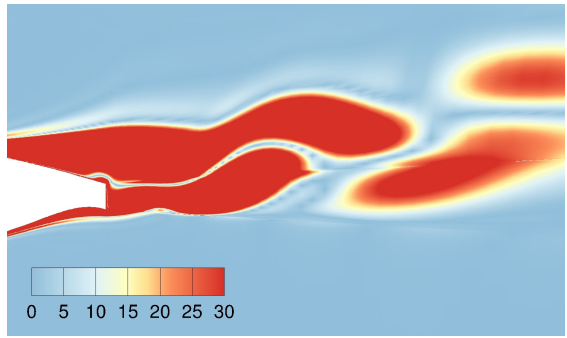

(b)

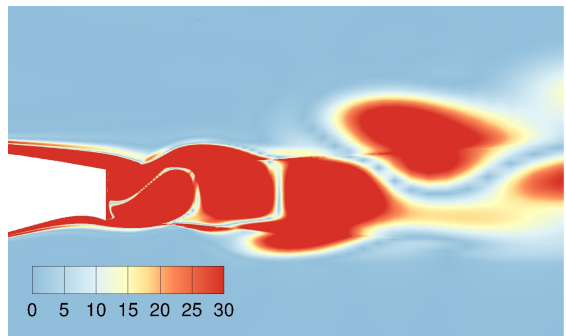

(c)

Fig. 5 Instantaneous vorticity contour near the trailing edge at AoA 12° for (a) 0.5% TE thickness, (b) 8.75% TE thickness, and (c) 17.5% TE thickness.

3.2 Wake and Unsteady Characteristics
The development of unsteady wake for thick flatback airfoils is presented in Fig. 5. It can be seen in Fig. 5.a that the airfoil variant with small trailing edge thickness tends to show the superiority of suction side separation beyond stall, then the domination decreases as the trailing edge thickness increases. Fig. 5.b shows the vorticity contour of the 8.75% trailing edge thickness airfoil. It should be noted that the airfoil is also in stall, however, the suction side separation is reduced and followed by the trailing edge vortex dominance. The contribution between these vortices is more or less equal. As the trailing edge thickness increases to 17.5%, the presence of suction side separation is almost negligible and the eddy structures are mostly provided by the trailing edge flow. The extent of vortical structures in shearwise direction behind the airfoil is significantly reduced (see Fig. 5.a-c) and this can be explained as follows: the flow on the suction side is more stable due to the APG reduction, however, the wake flow becomes more unstable and precipitous breakdown of the vortices cannot be avoided. Further studies are necessary to deeply reveal the characteristics of the wake flow over flatback airfoils.

These phenomena may be the reason behind the increase of the $c_l$ amplitude ($\Delta c_l$) with the trailing edge as shown in Fig. 6. It can be seen that the trailing edge thickness increase can lead to the significant augmentation of $\Delta c_l$. The value of $\Delta c_l$ for FB-3500-1750 at AoA 12° (max. $c_l$) is even higher than for FB-3500-0005 at AoA 20° (beyond stall). Fig. 6 also shows that $\Delta c_l$ decreases with AoA before stall, and increases afterwards. The explanation behind this, as already mentioned above, is due to the trailing edge or suction side vortices dominance.

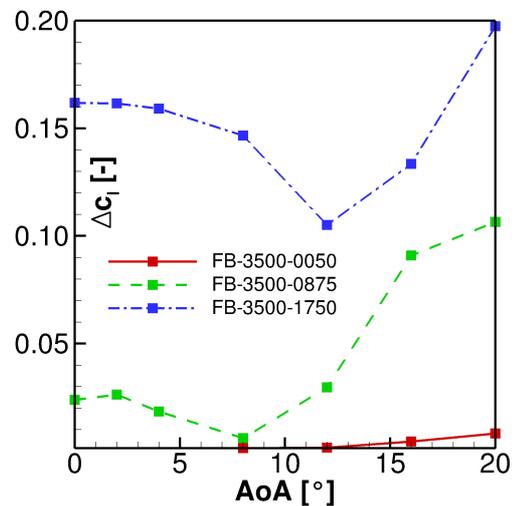

Fig. 6 Lift coefficient amplitude of flatback airfoils

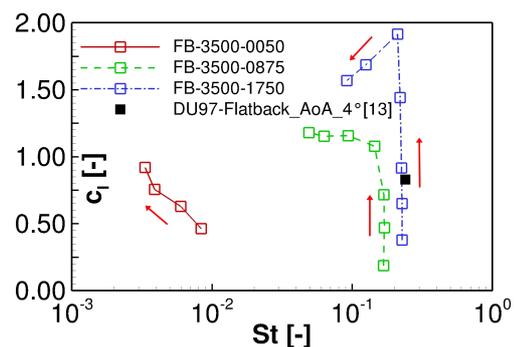

Fig. 7 Strouhal number of the lift coefficient (The arrow indicates the direction of local AoA increase).

The predicted Strouhal number (St) is calculated based on the lift coefficient frequency and it is nondimensionalized with respect to the freestream velocity $U_{inf}$ = 24.1m/s and the trailing edge thicknesses of 0.5%, 8.75%, and 17.5% chord length, with the chord value equal to 0.2032m. It is compared with measurement of the DU97-flatback airfoil [13] as shown in Fig. 7. The value of St before stall for FB-3500-1750 is very close to DU-97-flatback airfoil. The reason behind this is that they have similar ratio of trailing edge thickness to airfoil maximum thickness (δ). The Strouhal value increases with the trailing edge thickness and it is independent of the lift coefficient as well as AoA before stall, then it starts to decrease as the local AoA increases. It is expected that higher δ can lead to the higher value of Strouhal number.

The decrease of St as the airfoils operate at high AoA is mainly due to boundary layer separation. The separated shear layer produces larger vortices with lower shedding frequency, hence the Strouhal number decreases. These results are contrary to the works from Ref. [14] which found that the frequency of LRN(1)-1007 airfoil increases with the AoA. The difference is most probably due to the highly deviating airfoil thicknesses causing different stall mechanisms. This view has been supported by Ref. [15] which conducted measurements on a profile with 18% maximum thickness. The value of shedding frequency decreases with AoA, but its Strouhal number more or less remains constant since they defined the Strouhal number based on the vertical distance (chord*sin(AoA)) of airfoil (St*) [16]. This type of definition accommodates the separation occurring on the suction side of airfoil. However, it does not hold true with the existence of blunt trailing edge at low AoA. When the airfoil operates at zero AoA, St* is also zero since it has no vertical distance. It is obviously not appropriate for the present application to flatback airfoils. The use of boundary layer and wake length parameters for Strouhal number definition requires deeper examinations on the flowfield and is not directly applicable. Therefore, for blunt trailing edge, it is suggested to use the trailing edge thickness as the basis of St before the airfoil reaches stall and the vertical distance can be used afterwards.

## 4. Conclusion

The unsteady characteristics and flow phenomena of flatback airfoils have been discussed. The investigations confirmed that the use of flatback trailing edge increases the maximum achievable lift coefficient and delays the stall under tripped condition. It was shown that the unsteady characteristics in terms of Strouhal number as well as wake development are significantly dependent on the trailing edge thickness. A recommendation has been given in the definition of Strouhal number to express the unsteady parameter accurately in an applicable way.

## 5. Acknowledgement

The authors gratefully acknowledge Prof. C.P. van Dam and Sandia National Laboratories for provision of the airfoil coordinates, the High Performance Computing Center Stuttgart (HLRS) for the computational resources on LAKI, and Directorate General of Higher Education (DIKTI) - Indonesia for the scholarship grant.